\documentclass[10pt, conference,compsocconf]{IEEEtran}
\IEEEoverridecommandlockouts
\pdfoutput=1

\newcommand{\mtv}{metaverse}

\usepackage[hyphens]{url}
\usepackage{cite}
\usepackage{balance}
\usepackage{amsmath,amssymb,amsfonts}
\usepackage{algorithmic}
\usepackage{graphicx}
\usepackage{textcomp}
\usepackage{xcolor}
\def\BibTeX{{\rm B\kern-.05em{\sc i\kern-.025em b}\kern-.08em
    T\kern-.1667em\lower.7ex\hbox{E}\kern-.125emX}}
\usepackage{soul}

\begin{document}

\title{Metaverse: Security and Privacy Issues\thanks{This is the author’s version of the work. It is posted here for your personal use. Not for redistribution. The definitive Version of Record was published in IEEE TPS, December 12–15, 2021.}}

\author{
\IEEEauthorblockN{Roberto Di Pietro}
\IEEEauthorblockA{College of Science and Engineering (CSE) \\
Hamad Bin Khalifa University (HBKU) \\
Doha, Qatar \\
rdipietro@hbku.edu.qa}
\and
\IEEEauthorblockN{Stefano Cresci}
\IEEEauthorblockA{Institute of Informatics and Telematics (IIT) \\
National Research Council (CNR) \\
Pisa, Italy \\
stefano.cresci@iit.cnr.it}
}

\maketitle

\begin{abstract}
The metaverse promises a host of bright opportunities for business, economics, and society. Though, a number of critical aspects are still to be considered and the analysis of their impact is almost non-existent. 
In this paper, we provide several contributions. We start by analysing the foundations of the \mbox{\mtv}, later we focus on the novel privacy and security issues introduced by this new paradigm, and finally we  broaden the scope of the contribution highlighting some of the far-reaching yet logical implications of the \mbox{\mtv} on a number of domains, not all of them in tech. Throughout the paper, we also discuss possible research directions. \\*
We believe that the provided holistic view on the foundations, technology, and issues related to the \mbox{\mtv}---with a focus on security and privacy---, other than being an interesting contribution on its own,
could also pave the way for a few multidisciplinary research avenues.
\end{abstract}

\begin{IEEEkeywords}
Metaverse, metamedia, multiverse, singularity, security, privacy, Facebook, Meta
\end{IEEEkeywords}

\IEEEpeerreviewmaketitle

\section{Introduction}
\label{sec:introduction}
A metaverse is a combination of persistent, multi-user, shared, 3D virtual spaces that are intertwined with the physical world and merged together to create a unified and perpetual virtual universe. Users enter the metaverse with avatars, who can interact with each other and with the items, applications, services, and businesses that the metaverse contains. Originally tributed to the vision of  the American writer and tech advisor Neal Stephenson in his 1992 science fiction book \textit{Snow Crash}---but practically being an idea that can be found in many cultures and epochs, tracing back at least to {The Cave} of Plato \cite{the_cave}, as brilliantly described by Ethan Zuckerman in a recent issue of The Athlantic\footnote{\url{https://www.theatlantic.com/technology/archive/2021/10/facebook-metaverse-was-always-terrible/620546/}}---metaverses have periodically sparked heated debates among both tech experts and the general public, following more-or-less successful attempts at implementing them (e.g., Second Life). Recently, they have made the news again, when some of the biggest tech firms announced heavy investments and ambitious plans for the development of new and futuristic metaverses. Among them are
Microsoft\footnote{\url{https://twitter.com/satyanadella/status/1455624165201887234}}, Epic Games,\footnote{\url{https://www.epicgames.com/site/en-US/news/announcing-a-1-billion-funding-round-to-support-epics-long-term-vision-for-the-metaverse}} and Meta\footnote{\url{https://www.facebook.com/watch/live/?ref=watch_permalink&v=561535698440683}}---the tech holding in which Facebook was (uncoincidentally) rebranded. Mark Zuckerberg (Meta and Facebook CEO), in particular, envisioned the metaverse as the next evolution of the Internet. That is, a technology-empowered, cyber-physical Internet 3.0 capable of surpassing the mobile Internet paradigm. His vision of the metaverse as the next evolutionary leap of both our physical and digital networking capabilities---and thus, also of our social life---perfectly explains the massive hype and the spasmodic attention that rapidly built around the metaverse.

To support this ambitious vision, the first instances of the metaverse will build upon several recent technological advancements. For instance, virtual reality (VR) will be used to create immersive 3D spaces while augmented reality (AR) will allow for a tight integration between the virtual and the physical worlds. Along the same line, digital twins will allow physical objects to be brought, visualized, and shared into the metaverse. Wearable sensors will allow avatars in the virtual spaces to mimic real-world movements, while other sensors, such as those embedded into the next-generation of smart devices, will feed additional real-world data into the metaverse. The metaverse will also feature a rich marketplace of both physical goods and virtual items. The latter will be bound to, and owned by, the avatars themselves and will be implemented as non-fungible tokens (NFTs). Finally, the next generation of networking technologies and algorithms will make the metaverse even more pervasive than current social media and social networking platforms. 

While incorporating the latest technologies, social media, advanced algorithms, a constant and boundless supply of data from the developing sensory systems, and much more, the \mtv\ will be still more comprehensive than what could be described at the time of writing, because of its capacity of adapting and adopting any innovation. In this context, it appears evident that security and privacy concerns will not only be magnified, 
but also completely transformed. 
And with them, a few  other critical issues, as discussed in the following.

\subsection{Contributions}
In this paper we provide several contributions. We first revise the genesis of the concept of \mtv. Later, we describe its current background, which helps us explain why the \mtv\ is a hot topic just now, despite having being around for ages. We highlight current security and privacy issues, which in this new context are magnified as they have never been before. We further highlight why, this time, the context is radically different from what we experienced in the past: that is, why the upcoming \mtv\ instances could represent something semantically different from what we have seen in the technology domain until now (an approximated answer: exponential technology growth; unprecedented capabilities in data generation, collection, and analysis;  and, convergence of the cyber-physical worlds). Later, we seed the concept that we use to forecast the development of the \mtv\ in the years to come. Based on the introduced foundational material we develop our vision of the \mtv, highlighting the related risks for security and privacy.
Finally, in our discussion we argument the thesis that   the exposed threats and the envisaged impacts generated by the development of the \mtv\ call for an interdisciplinary approach where computer science and engineering are critical, but at least to the same extent as other apparently far disciplines, like philosophy, law, and social sciences.

\subsection{Roadmap}
The remainder of the paper is organized as follows. In Section \ref{sec:background} we revise some background information and related work in the field. In Section \ref{sec:issues}, we discuss the current issues of \mtv, with a specific reference to security and privacy. 
In Section \ref{sec:discussion}, based on the state of play and the introduced guiding principles, we provide our analysis of what are the most stringent criticalities of the \mtv, also highlighting broad---in scope, depth, and breadth---research directions.
Finally, in Section \ref{sec:conclusions} we provide our conclusions.

\section{Background and related work}
\label{sec:background}
Many of the technologies that represent the building blocks of the metaverse already exist at an advanced state of development. However, many others are still several years away from being usable in practice. In the remainder of this section we provide an introduction to the metaverse by describing its core elements, characteristics, and functionalities, as well as the long term directions in which it could further expand. Without loss of generality, we take as a starting point---and later develop---the vision of the \mtv\ proposed by Zuckerberg, since it represents the most comprehensive one among all those that have been proposed so far. 

\subsection{Core Elements, Characteristics, and Functionalities}

\subsubsection{Activities}
The metaverse is a virtual universe, or a substrate, capable of supporting and interconnecting a multitude of different applications. As such, the activities that users can carry out in the metaverse are as diverse as the applications embedded in it.

The unprecedented networking opportunities enabled by the metaverse make it particularly convenient for engaging in \textit{social} activities. Traditional activities such as befriending other users, or engaging in chats and audio/video calls will be supported in the metaverse too. One way in which these functionalities will be made available is by integrating existing messaging and videoconferencing apps into the metaverse. In addition to these activities, which barely represent a porting of already-existing interaction schemes, the shared virtual spaces of the metaverse will also enable additional forms of social interaction---for instance, the interactions between 3D avatars that are typical of massively multiplayer online games (MMOs). Regarding the latter, \textit{gaming} and other forms of \textit{entertainment}, including the possibility to participate to art shows and concerts, will represent another major group of metaverse activities. Firstly, as previously anticipated, metaverses inherit several characteristics from MMOs. In addition, the gaming sector is in constant growth, both in terms of revenue and users. The combination of these two factors ensures that gaming, and entertainment in general, will be among the most frequent activities in the metaverse. Notably, metaverse shows can be both natively virtual, as in the case of the many concerts held within the virtual worlds of online games such as Fortnite, Minecraft, and Roblox\footnote{\url{https://www.nme.com/features/gaming-features/fortnite-roblox-best-in-game-concerts-2021-3021418}}, or natively physical but nonetheless accessible via the metaverse, such as in the case of a real-world concert that allows metaverse users to participate via VR. \textit{Sports and fitness} are another group of activities that will benefit from the cyber-physical integration enabled by the metaverse. In particular, wearable sensors and AR/VR will allow for realistic and immersive virtual sport simulations, with unprecedented opportunities of personalization and customization. The same considerations can be made for \textit{learning} and other \textit{educational} activities, which will greatly benefit from the immersiveness and 3D capabilities of the metaverse. Finally, the metaverse will also be used for \textit{work and business}, as well as for \textit{commerce}. Regarding the former, digital twins, VR, and the availability of embedded messaging and videoconferencing apps will allow rich, immersive meetings to take place in the metaverse. In addition, traditional and new forms of commerce will be supported by one or more online marketplaces, which will feature both physical and digital goods for sale. About the latter, in particular, the marketplace will connect independent content creators with their potential customers (i.e., metaverse users), allowing business opportunities to scale to unprecedented levels.

\subsubsection{Immersiveness}
In the past decade, we witnessed to the revolution of the mobile Internet, enabling  access to our online applications and social ecosystems from virtually everywhere. Currently however, our capacity of entering and enjoying virtual environments is constrained by the use of screens and mobile devices. With the advent of the metaverse, instead, access to online virtual spaces will be possible also via AR and VR technologies. Indeed, two of the key and distinguishing features of the metaverse are its pervasiveness and immersiveness, reached via an unprecedented merge between the virtual and the physical worlds. The metaverse is pervasive and immersive in multiple ways, involving both the ways in which we access and interact with it, as well as the ways in which we receive feedback from it. For instance, mobile immersive access to the metaverse will be possible thanks to the next generation of AR-enabled smart devices (e.g., compact smart glasses\footnote{\url{https://www.theguardian.com/technology/2021/sep/15/techscape-smart-glasses-facebook}}). Instead, work or home access will be possible with lightweight and comfortable VR goggles. The switch from 2D interfaces to 3D virtual spaces will be accompanied by a number of additional possibilities that will increasingly boost immersiveness. Firstly, many of the 2D applications and services that we use on a daily basis (e.g., Dropbox, Slack, Zoom, Facebook, Instagram, and many more), will become applications embedded into the metaverse. Then, users will inhabit the metaverse in the form of avatars, thus switching from static 2D profile images to interactive and personalized 3D avatars. Depending on the activity, application, or the virtual space in use, users will be able to represent themselves with either photorealistic, cartoonish, or fully fictional avatars. Users will also have the possibility to create virtual copies of physical items (i.e., digital twins) and to share them in the metaverse, thus further reducing the gap between the virtual and physical dimension. Finally, the use of wearable sensors and devices will tighten the bond between our physical and virtual worlds by feeding orders of magnitude more real-world data into the metaverse and by giving users unprecedented sensory feedback~\cite{haptic}.

\subsubsection{Interoperability}
From an architectural standpoint, the metaverse can be regarded as a unifying framework or substrate that connects the multitude of applications and services that are embedded into it. As such, interoperability is another key feature of the metaverse and its users will experience it in multiple ways. For instance, they will be able to interact simultaneously with multiple applications, similarly to what we usually do in our desktop computers or mobile devices. Today, this level of interoperability between different apps is ordinary and expected for general purpose physical devices. However, it is unprecedented for online virtual environments. Think for instance to MMOs--massive virtual worlds where players can typically perform only a limited subset of similar and related activities. Extending this concept, in the metaverse also spaces and activities will be interconnected. In fact, it will be possible to seamlessly move across different virtual thematic spaces, or to interrupt an activity in order to start a new one (e.g., stopping a game set in a dedicated space in order to join a friend in another space). Virtual items, such as avatar outfits, will also be part of this interconnectedness. Indeed, in one of the possible evolutions of the \mtv, items will be owned by the users, instead of the platforms, and the interoperability of the metaverse will allow users to buy certain virtual items as NFTs from an applcation's store and to use them with their avatars in other applications and spaces, and throughout all of the metaverse.

\subsection{Beyond Current Technology}
The description of the metaverse that we provided so far, its uses and activities, as well as its key elements and characteristics, is based on existing and relatively established technology. Hence, the development of a metaverse that supports this vision is expected to occur in the short-medium term. At the same time, however, metaverses promise to deliver other additional and pioneering functionalities---extending by far the possibilities that are enabled by  the current state of technology. These latter functionalities inevitably depend on the scientific and engineering progress that is expected to be achieved in certain technological areas in the coming years, and will possibly be available in the metaverse only in the medium-long term. Nevertheless, the analysis of these advanced functionalities is instrumental for understanding the full extent of opportunities and challenges resulting from the metaverse.

An area where rapid technological progress is expected is the one related to the interfaces that will allow interacting with the metaverse. These include the technologies and the devices used for inputting commands to, as well as for receiving feedback from, the metaverse. Among them are brain–computer interfaces (BCIs)---that is, neural interfaces designed to collect and process the electrical signals generated in the human brain as a result of some cognitive activity, and to convert them into meaningful inputs for an external computer or apparatus~\cite{zhang2020electronic}. Notably, such interfaces can also be used to influence the brain, as in the case of those designed to treat depression and other mental disorders. In addition to the existing noninvasive BCIs, the US military and some tech firms---including Meta and Elon Musk’s Neuralink---are also researching more powerful interfaces based on electrodes implanted directly into the brain~\cite{horgan2021should}. On the one hand, BCIs hold the potential to revolutionize the way in which we input commands to machines, freeing us from the constraints derived from the use of mouse \& keyboard---a particularly desirable feature to have during a hectic VR session. On the other hand,  they also open new possibilities for delivering powerful (neural) feedback---another area where much scientific and technological progress is expected. In fact, the metaverse will eventually involve multi-sensory experiences and feedback, be it by means of brain implants, or via other technologies (e.g., haptic devices)~\cite{haptic}. For instance, sensory systems will provide metaverse users force-return effects that mimic the physical interactions in the real world, depending on the outcome of their actions in the 3D virtual world. To this regard, the metaverse will represent the next evolutionary step in our capacity to deliver and consume not only multimedia, but also multi-sensory content---a logical evolution with respect to what we have already experienced. Indeed, the Internet was originally conceived as a technology for exchanging text messages (e.g., emails). Subsequently, with the advent of the social Web, we witnessed to the spread of multimedia content in the form of images (e.g., memes) and videos. Eventually, the Internet 3.0 will be a multi-sensory experience. It is precisely this outlook on the metaverse as the application that will lead us into the era of Internet 3.0, that motivates some scholars to refer to it also as a \textit{mediaverse} or \textit{multiverse}~\cite{woodgate2021metaverse}.
 \section{Current Issues}
\label{sec:issues}
In the previous sections we briefly sketched the main functionalities of the metaverse---both in its likely initial implementations, as well as in the more advanced ones---and we highlighted the plethora of new possibilities opened up by this novel socio-technical paradigm. However, this unprecedented degree of immersiveness and interoperability also has a flip side: an increase in quantity and quality of the threats associated to the current technologies that will be adopted to realize the \mtv, accompanied by the generation of a few  equally-unprecedented threats. Such threats are especially related to the \textit{privacy} and \textit{security} of metaverse users, which we discuss in the remainder of this section.

\subsection{Privacy}
In nowadays Internet, it is said that if you do not pay for a product or service, then \textit{you} (or rather \textit{your data}) are the product. Social media and social networking platforms are the paramount example of this kind. These platforms offer {\em free} services that involve millions or even billions of users, whose preferences are so well-known to the platforms themselves~\cite{kosinski2013private} that they are able to show users extremely accurate, micro-targeted advertisements. This successful business model is possible only because of the platform's capability to accurately profile its users, by analyzing their actions and interactions with the platform's content and with the other users, other than relying on further worrisome tracking capabilities, thanks to the evolution of cookies and, generally,  fingerprinting techniques \cite{browse_fingerprinting}. Even with today's technology, the digital crumbles that we leave behind us already tell a lot about our personality, tastes, and orientations (e.g., political and sexual). This was evident since the very first studies 
carried out almost one decade ago \cite{political_twitter,kosinski2013private}, while these days such predictive capabilities have developed  exponentially. Given these assumptions, what could happen in the \mtv ? In the following subsection we strive to imagine the data collection capabilities, and related applications, enabled by the \mtv.

\subsubsection{User profiling in the \mtv}
If social network users are the product of today's Internet, in the metaverse literally \textit{everything and everyone} will be the product. Social networking platforms currently act as powerful magnets for Web users. Similarly, the metaverse will be an exponentially more powerful magnet for (even more) users, as well as for content creators, entrepreneurs, and businesses alike. In other words, it will be a unified meta-platform for users---independently from their passions and preferred applications (e.g., readers, gamers, students, etc.)---as well as for the developers of such applications and the businesses that run them. 
The exposed consideration  raises major concerns over the amount and type of data that such a massive platform could collect. The internet 2.0 allowed marketers to study where users move their mouse, where they look on a screen, how much time they spend on a given pictured item, and which products or users they like. At times, and especially for technology illiterate users, one is not even aware that such recordings and analyses take place and, hence, their privacy may be in jeopardy in unanticipated ways~\cite{falchuk2018social}. In the metaverse, current data collection techniques and related analyses will be considered amateurish, at best. Indeed, the platform will be able to track our body movements,  physiological responses,  likely even brainwaves, and real and virtual interactions with the surrounding environment, to cite a few. Moreover, these capabilities will be in addition to all other data that are already being collected.\\* How will such data be used and what are the risks for user privacy? In the following, we try to address the latter question.

\subsubsection{User privacy}
Regarding user privacy in the metaverse, three areas are particularly relevant~\cite{falchuk2018social}: (i) personal information; (ii) behavior; and, (iii) communications. As a result of our previous considerations, each of these areas will give to the platforms much more data than what they currently have, with new and increased risks. As an example, personal information collected from social networking platforms are already used for \textit{doxing}---that is, the practice, or the menace, of revealing private information of a victim with the aim of extortion or for online shaming~\cite{snyder2017fifteen}. Given that the metaverse will provide much more personal information about its users, not only to the platforms, but also to other users, how will we keep doxing at bay? Notably, personal and sensitive information that will leak through the metaverse will include a plethora of real-world information about user habits and their physiological characteristics. While these are difficult to obtain in the current Internet, if not outright impossible,  they will be much more easily acquired in the metaverse, as a result of the tighter bond between the virtual and physical worlds.

This leads us to the risks related to the privacy of user behavior. To this regard, the metaverse will offer unprecedented opportunities for exploiting online immersive experiences and interactions to perpetrate offline (i.e., real-world) harms and frauds. Indeed, social engineering attacks already account for the largest share of cyber-attacks suffered online~\cite{salahdine2019social}, as also measured during the COVID-19 pandemic. With the metaverse, social engineering attacks will likely become even more convenient and powerful, and thus, more frequent. In addition to social engineering, the metaverse raises additional concerns related to the privacy of user behaviors. \textit{Spying} and \textit{stalking} are practical examples of this kind. In the real-world, eyesdropping, following, or harassing someone can be partially hindered by physical constraints, such as the need to be physically close to another person and to move to certain locations, which might also involve some  cost (e.g. time, money). Notably, the cited penalties generally act as  excellent deterrents. However, the same considerations do not hold in the metaverse, which makes such attacks more convenient. Unfortunately, this already applies to a large array of attacks that currently proliferate online, some of which are often perpetrated by multiple coordinated users, and that will likely skyrocket in the metaverse~\cite{cresci2017social,weber2021amplifying}. Among them are coordinated harassment and raiding, shaming, cyberbullying, video call bombing, and shitstorming~\cite{ling2021first,flores2018mobilizing}, to cite a few. Some of these behaviors have already been used as forms of ``denial of service''. For instance, in online games---which will be one of the primary uses of the metaverse---a few toxic gamers are enough to repeatedly ruin the game for all other participants~\cite{bakioglu2009spectacular}. Moreover, many cyber-aggressions that initially start on a specific platform or in relation to a specific topic (e.g., a game) can also subsequently expand to other platforms or topics, thus involving additional users and communities, as it happened in the case of the \#Gamergate campaign~\cite{chatzakou2017measuring}, or even in Second Life~\cite{falchuk2018social,bakioglu2009spectacular}. In a metaverse characterized by a multitude of interconnections between communities, spaces, and applications, these risks are inevitably amplified.

Finally, more connections imply more interpersonal communications, which leads to an increase in the number and manners in which information could be collected and misused, and cyber-crimes could be perpetrated. Privacy concerns about metaverse communications are not restricted to the obvious risks of corporate data breaches, but also involve other forms of communications between users. Think for example to sexting---the practice of exchanging sexually explicit messages---, which is currently carried out via 
mobile phones~\cite{geeng2020usable}. Sexting, or other forms of sexually-oriented communications and interactions, could become common in the metaverse, also thanks to its rich and multi-sensory 3D world~\cite{bardzell2007docile}. What if the privacy of such personal communications is endangered? In the Internet 2.0, revenge porn---that is, the distribution of sexually explicit texts, images, or videos of individuals without their consent---is largely confined to certain not-safe-for-work (NSFW) platforms. Similarly, toxic users are clustered in fringe Web platforms or in largely isolated communities of like-minded peers~\cite{zannettou2018origins}. Disgruntled employees have relatively few ways for publicly harming their company's reputation in a scalable way. However, each of these crucial---yet, so far, fringe---issues could become mainstream in the massively interconnected metaverse.\\*
What can be done to address the above exposed threats? A preliminary discussion is provided in the following subsection.

\subsubsection{Countermeasures}
Given the multitude of privacy risks to which metaverse users are exposed to, some scholars already started envisioning ways in which to enforce user privacy in 3D social metaverses~\cite{lee2021all}. Among them, a few solutions have been proposed~\cite{falchuk2018social}, that are  based on the combination of three fundamental strategies: (i) creating a mannequin or multiple clones of one's avatar to shadow one's own activities; (ii) creating a private copy (e.g., an instance) of a public space for the exclusive use of the user, or temporarily locking out other users from a public space; and, (iii) allowing user teleportation, invisibility or other forms of disguise. Meaningful combinations of the above strategies can also be used, such as applying a disguise to one's avatar after exiting from an instanced space, so as to avoid being recognized and chased. 
Independently of the privacy solutions that a metaverse implements, these ones should be made available to the users (e.g., via a privacy menu) so that they could choose their desired level of privacy, also depending on their activities, and the way in which to apply the selected privacy features. However, all the (few) solutions envisioned so far are designed for simplistic metaverses, those that have existed until now. 
As such, they are much likely not sufficient to withstand the risks and attacks of complex, immersive and massively interconnected multiverses, such as the one envisioned by Zuckerberg's Meta. To this regard, new and better solutions should be devised---an exciting and daunting research challenge.

\subsection{Security}
The innovative combination of the multiple powerful technologies that will come into play in the \mtv\ will spark a torrent of security threats, in addition to the privacy issues already discussed. A few of them are highlighted in the following. 

\subsubsection{Humans in and out of the loop}
Some of these security concerns derive from the inherent complexity of the metaverse. In fact, it is easy to envision that a meta-platform connecting orders of magnitude more users, services, applications and goods---and ultimately, handling much more data---than those managed by the current Web platforms will inevitably require metaverse administrators to push automation, that is  to tackle more tasks with \textit{algorithms}, rather than with human operators. The need to delegate tasks and operations to algorithms, especially to those implemented with cutting-edge AI techniques, emerges both from the need to achieve great scalability as well as efficiency and performance. However, in the current version of the Internet we are starting to understand the dire implications of delegating societally relevant tasks to algorithms that, in spite of achieving unmatched performance, are affected by several issues. Among them are biases that can prevent fair outcomes~\cite{corbett2017algorithmic}, the lack of transparency~\cite{datta2016algorithmic}, the vulnerability to attacks and manipulations~\cite{mcdaniel2016machine,cresci2021adversarial}, and the huge computational and energy requirements of complex AI models (e.g., deep learning), which limit both their affordability\footnote{\url{https://towardsdatascience.com/the-future-of-computation-for-machine-learning-and-data-science-fad7062bc27d}} and sustainability\footnote{\url{https://www.forbes.com/sites/glenngow/2020/08/21/environmental-sustainability-and-ai}}. Each of these issues represents an open scientific challenge that will require many conjoint efforts from different scientific communities (e.g., AI and machine learning, security, ethics, and more) to be tamed and solved. 
In the meantime however, our reliance on ``problematic'' algorithms is leading to worrisome problems. Moreover, always more frequently, problems that originate online have offline consequences, as in the notable cases of the January 2021 Capitol Hill riots that resulted from the dramatic polarization of our online social environments, or the rampaging COVID-19 and vaccine misinformation that costed many human lives~\cite{dipietro2021new,ferrara2020misinformation}. In the metaverse---a much bigger and massively more interconnected platform---which new problems will arise? More importantly, are we ready to delegate so much of our lives to the algorithms in the metaverse? 
In addition to the exposed open questions, it is worth noting that overcoming the security challenges posed by the algorithms exploited in the metaverse will require the design and development of new methodologies and technical solutions, as well as legal ones (e.g., who would be responsible for the mistakes made by an algorithm? Would cost externalization be possible?).

\subsubsection{Integrity and authentication in the metaverse}
As a practical and relevant example of the problems caused by algorithms and automation, we can cite the security issues of content \textit{integrity} and user \textit{authentication}. Even in nowadays' social networking platforms, a large share of our interactions occur with inorganic or fabricated content (e.g., machine-generated text) and with inauthentic users (e.g., human-operated fake personas that troll other users, or even fully automated accounts such as bots). For instance, regarding software-driven accounts, it has already been shown that humans can be ``fingerprinted'' and digitally reproduced in social media, without other humans nor detection algorithms noticing it~\cite{cresci2020decade}. Moreover, it has also already been envisioned that future advances in AI will make such automated accounts totally indistinguishable from humans~\cite{boneh2019relevant}. In the metaverse, these human-machine interactions will become more frequent and, at least for certain activities, even mandatory. Think for instance to the need of resorting to chatbots and in-game bots, in order to deliver a rich, immersive, and credible user experience~\cite{falchuk2018social}. These capabilities raise  new and additional security concerns. For example, how easy will it be to fake one's own age, gender, job, or any other attribute? Will we be able to tell humans and machines apart? How will human or platform vulnerabilities be exploited by automated accounts and machine-generated content, in such a complex, immersive and persuasive world? Earlier attempts at exploiting the vulnerabilities of contemporary platforms already managed to shake the foundations of our democratic societies. The next wave of attempts might prove fatal.

\subsubsection{Polarization and radicalization in a singleton world}
While the previous security issues emerged from the necessity to rely on algorithms and automation for managing a huge virtual world, other problems will surface as a consequence of the uniqueness of the metaverse---that is, its characteristic of being a \textit{singleton}. Indeed, a metaverse is essentially a massive aggregator of applications, services, goods and thus, also of users. As all platforms that collect, aggregate, and deliver content, its success will depend on its capacity to act as a centralized access point for such contents. Notably, this characteristic is in sharp contrast with the current structure of the Web, which features a multitude of different platforms, some of which are mainstream and humongous while others are fringe and minuscule. Put differently, the plurality of the existing Web platforms will be replaced by, at most, a handful of massive metaverses. The plurality of the Web also means that each user, independently on her tastes and preferences, will likely find an online platform or a community of users with the same tastes and preferences. This is the reason why fringe and minuscule Web platforms exist in the first place: in spite of their overall marginality, they serve a useful purpose (at least for those who browse them). In contrast, a singleton metaverse will force the simultaneous presence, coexistence, and interaction of all users, including those with peculiar interests and opposed viewpoints, as well as those that would have never got in touch on the Web. While, to a certain extent, this already happens in the current version of the Internet, the relatively limited interconnectedness and the filter bubbles of today's virtual spaces (e.g., platforms and applications) ensure that different types of users mainly belong to different online communities, possibly even residing on completely autonomous and independent platforms. However, when radically different---possibly even opposing---groups of users merge, the consequences can be severe. For instance, we know from massively multiplayer online role playing games (MMORPGs) that certain types of players tend to systematically harass other types, as in the case of male, low-skilled players with regards to female players~\cite{kasumovic2015insights}. Hence, what would happen when all sorts of different communities merge in a centralized, shared metaverse? To a certain extent, each user that populates an online space, such as today's social networks and the upcoming metaverse, has her own interests and ways of enjoying it. In other words, each user behaves in a different way. Some users may thus exploit their way of ``behaving'' in the metaverse to troll, harass, or anyway to take advantage of other users, in new and unanticipated ways.
 \section{Discussion}
\label{sec:discussion}
In order to be able to understand the implications of the \mtv,  we believe that it is useful to refresh the concept of ``singularity'' and to analyze its roots. From there, we will move forward in providing  support to thesis that  the  \mtv\ is where the singularity will take place and, finally, we will sketch some of the possible implications.

The singularity can be described as a point in time at which technological growth becomes uncontrollable and irreversible, resulting in unforeseeable changes to human civilization. This concept is particularly compelling in its reformulation applied to AI: ``The AI singularity refers to an event where the AIs in our lives either become self-aware, or reach an ability for continuous improvement so powerful that it will evolve beyond our control''\footnote{\url{https://www.singularitysymposium.com/definition-of-singularity.html}}. Not surprisingly, the original concept of singularity is attributed to John von Neumann. However, one might notice that the concept of singularity could be (almost equivalently) rephrased and generalized as: ``the gradual accumulation of quantitative changes that would, at some point (at break measures), turns things into another or a new quality, entailing new and quantitative characteristics''. If the latter formulation sounds familiar, it is because that formulation of singularity is actually the second law of dialectics by the 18th century's German philosopher G.W.F. Hegel\footnote{It is likely that Van Neuman knew the laws of dialectics, but at that time, a direct reference to Hegel would have probably been not well accepted (McCarthyism was ravaging the US), and hence the---brilliant!---re-formulation and adaptation of the cited second law of dialectics into ``singularity'', that we will use as well in the remainder.} \cite{Hegel1929_SoL}. \\*
What we just described provides the foundations for the reasoning on the evolution and implications of \mtv.

Indeed, it seems quite reasonable that the convergence of technology, platforms, and sensory systems (to cite a few) into a single domain---the \mtv---, combined with the sheer amount of data that will be generated therein, do qualify the \mtv\ as  the best candidate for a qualitative transformation of the very same domain. If this hypothesis is true, we need a paradigm shift in our analysis tools to deal with the \mtv. 
Indeed, if  we keep  analyzing  the \mtv\ following the usual approach adopted in hard sciences---that is, by considering the single components and by studying the corresponding interactions of those components---the results of our analysis  would be probably  wrong, or at least incomplete. Indeed, the \mtv\ would not equal the sum of the single, different components used to structure it. The \mtv---the developed version of it, not the approximation that we will see in the next few years---will be something unique, never seen before. Though, the very seeds of it are being planted now, and the next few years will likely represent the most precious (and maybe the only) time window of opportunity where, as a community, we could still intervene to shape its strategic development directions.

In particular, if we want to influence the development of the metaverse and, especially, try to anticipate its evolution so as to be able to figure out what are the novel risks introduced by it, we will need a more comprehensive approach. Simply put, we would need the best minds, supported by the whole arsenal of methodologies, approaches, and tools developed in a range of disciplines, ranging from  Computer Science and Engineering, to  philosophy,  law, and, psychology, to cite a few.  This would be probably the only way to confront with the singularity that the \mtv\ will likely generate. Otherwise---that is, if we do not anticipate the complexity of the \mtv, or if we fail to forecast its full implications and their possible impact---the incurred risks would be staggering. In particular, we present in the following the most likely one. That is, we will embrace the \mtv\ with a false sense of control----being able to keep under control just some specific aspects of it, while completely ignoring the novel elements generated by the singularity. In such  a scenario, there would be no major transformations  for a short while, and the \mtv\ will even provide possible short-term benefits for users, owners, and the society. However, there is an impending criticality to this short-sighted approach: without a clear comprehension of the long-term implications of the \mtv\ on the several dimensions it will impact (social, economical, political, anthropological, etc.), there is a high chance that the \mtv\ will be eventually misused in manners and to an extent that are difficult to predict. To make an example, and not even a dreadful one, 
the Cambridge Analytica scandal~\cite{chen2018cambridge} (and its dire effects) would be considered nothing more than a childish game.

Having provided (re-discovered) a generalization for the concept of singularity---a key concept to capture the essence of the \mtv---, having clarified why we expect a convergence of different technologies and platforms, and having sketched the possible implications of the directions the \mtv\ could take, we can now leverage the privilege conceded to a vision paper to draft, in the next section, some concluding remarks that, we are afraid, will raise more questions  rather than highlighting solutions.

 \section{Conclusions}
\label{sec:conclusions}
The \mtv\ is approaching fast. Not because a of a communication campaign organized by a tech and social media behemoth to escape public scrutiny\footnote{\url{https://www.nytimes.com/2021/10/29/technology/meta-facebook-zuckerberg.html}}, or because it is just  a big  business opportunity. It is coming true because we are at the beginning of the singularity. Or, in simpler terms, because the time is ripe. The exponential progress of technology has brought us miniaturized sensing devices with the computing and communication capabilities of laptops and has spread the adoption of technology to all the aspects of our lives (e.g., economics, politics, industry, social relationships, etc.).  In addition, the social acceptance of an ever increasing rate of data recording and sharing of our personal experiences has provided the bulk of data to make the \mtv\ attractive, while  
5G and the coming 6G will resolve the  remaining gap in connectivity and sensing. 
On top of that, machine learning and AI algorithms are already efficient enough to predict with surprisingly good approximation our needs and actions, and keep improving. 

The (partial) list of cited technologies and paradigms, taken one by one, could represent a leap forward---possibly even a giant one---but just a leap forward on a path that is already known. Instead, our thesis is that the combination of the cited forces would produce an effect that is magnified not only quantitatively, but also qualitatively, thus dramatically and permanently changing the tech and the cognitive landscapes. In this new world that we are starting to  call \mtv, the traditional systemic threats imported by the technology, like security and, especially, privacy are critical ones. However,  in the \mtv~these threats could present themselves  in a manner that would (at least partially) escape the logical and technological schemes we have developed so far to cope with them. But, what is most, it may well be the case that security and privacy threats will not even represent the most critical ones. As we have substantiated in this paper, the \mtv\ arises complex issues, that just partially pertain to the technical domain. That is why the only way to navigate this {\em incognito} universe we are starting exploring, is to study it via a truly multidisciplinary and broad approach that conjugates technical fields with humanities and social sciences. To let the \mtv\ become a dream, or either discover that it has turned into a nightmare.
 
\balance
\bibliographystyle{IEEEtran}
\bibliography{references}

\end{document}